\documentclass[a4paper,11pt]{article}

\usepackage{setspace}
\usepackage{color}
\usepackage{pdfcolmk}	% so that colored text is continued after a page break with pdftex
\usepackage{mathptm} 
\usepackage{graphicx}
\usepackage{ulem}

\newcommand{\degC}{{\,}^\circ C}

\def\corr#1{#1}				% to get rid of corrections

\def\preprint#1{ }		% compressed mode

\preprint{ \large\normalsize}

\begin{document}

%---------------------- TITLE PAGE ---------------------------------
% The title page must contain the title of the article; author(s)
% name(s); all departments and institutions in which the work was done;
% an abbreviated title for the running head; and the name, e-mail, and
% address for correspondence.

\title{Calculating event-triggered average synaptic conductances \\
from the membrane potential}

\author{Martin Pospischil, Zuzanna Piwkowska, Michelle Rudolph,
Thierry Bal and Alain Destexhe$^*$ \\ UNIC, CNRS, Gif-sur-Yvette,
France \\ \ \\ $*$ Corresponding author (destexhe@iaf.cnrs-gif.fr) \\
\ \\ Abbreviated title: ``Spike-triggered average conductances from
Vm activity'' \\ \ \\ {\it Journal of Neurophysiology} \ {\bf 97}:
2544--2552 (2007)}

% Keywords: Event-triggered average, conductance-based models,
% integrate-and-fire,-clamp}

\maketitle

\thispagestyle{empty}	% no page number on title page

\preprint{\ \\ \noindent {\bf Address for correspondence:} \\ \ \\
Alain Destexhe, \\ Integrative and Computational Neuroscience Unit
(UNIC), \\ CNRS, 1 Avenue de la Terrasse (BAT 33), \\ 91198
Gif-sur-Yvette, France \\ Tel: +33 1 69 82 34 35 \\
destexhe@iaf.cnrs-gif.fr}

\preprint{\clearpage}

%---------------------- ABSTRACT ---------------------------------
% An informative one-paragraph abstract of not more than 250 words must
% accompany each manuscript. Note that longer abstracts are usually cut
% off at the end when displayed on Medline. It must state concisely
% what was done and why (including species and state of anesthesia),
% what was found (in terms of data, if space allows), and what was
% concluded.

\subsection*{Abstract}

The optimal patterns of synaptic conductances \corr{for} spike
generation in central neurons is a subject of considerable interest. 
Ideally, such conductance time courses should be extracted from
membrane potential (V$_m$) activity, but this is difficult because
the nonlinear contribution of conductances to the V$_m$ renders their
estimation from the membrane equation extremely sensitive.  We
outline here a solution to this problem based on a discretization of
the time axis.  This procedure can extract the time course of
excitatory and inhibitory conductances \corr{solely} from the
analysis of V$_m$ activity.  We test this method by calculating
spike-triggered averages of synaptic conductances using numerical
simulations of the integrate-and-fire model subject to colored
conductance noise.  The procedure was also tested successfully in
biological cortical neurons using conductance noise injected with
dynamic-clamp.  This method should allow \corr{the extraction of}
synaptic conductances from V$_m$ recordings in vivo.

\preprint{\clearpage}

%---------------------- INTRODUCTION -------------------------------
% Provide a brief overview of the scope and relevance of the study,
% especially with regard to previous advancements in related fields.

\subsection*{Introduction}

Determining the optimal features of stimuli \corr{which are needed to
obtain} a given response is of considerable interest, for example in
sensory physiology. Reverse-correlation is one of the
\corr{most-used} methods to obtain such estimates (Ag\"uera y Arcas
and Fairhall, 2003; Badel et al., 2006) and, in particular, the
spike-triggered average (STA) is often used to determine optimal
features linked to the genesis of action potentials (de Boer and
Kuypers, 1968).  The STA can be used to explore which feature of
stimulus space the neuron is sensitive to, or to identify modes that
contribute \corr{either} to spiking or to the period of silence
before the spike (Ag\"uera y Arcas and Fairhall, 2003).  Using
intracellular recordings, it is straightforward to calculate the STA
of the membrane potential (V$_m$), which yields the mean voltage
trajectory preceding spikes.  In contrast, it is much harder to
determine the underlying synaptic conductance.  Straightforward
methods like recording at several different DC levels and estimating
the total conductance from the ratio $\Delta I / \Delta V$ fail,
since the presence of a voltage threshold necessitates $\Delta V
\rightarrow 0$ at the time of the spike, \corr{which}, in turn,
artificially suggests a divergence of the total conductance to
infinity. Similarly, solving the membrane equation for excitatory and
inhibitory conductances separately suffers from an additional
complication: because the distance to threshold changes, the time
courses of the average synaptic conductances depend on the injected
current.  

Recent contributions (Badel et al., 2006; Paninski 2006a, 2006b) gave
analytical expressions for the most likely voltage path, which in the
low-noise limit approximates the STA, of the leaky integrate-and-fire
(IF) neuron.  In those cases, gaussian white noise current was
considered as input.  In Badel et al.\ (2006), a second state
variable was added in order to obtain biophysically more realistic
behavior.  In Paninski (2006a, 2006b), in addition the exact voltage
STA for the non-leaky IF neuron was computed, as well as the STA
input current in discrete time.  Here, a strong dependence of the STA
shape on the time resolution $dt$ was found without a stable limit as
$dt \rightarrow 0$.  It was argued heuristically that this behavior
results from the fact that decreasing the time step corresponds to
increasing the bandwidth of the input current, a point which was
supported by numerical simulations (Paninski et al., 2004; Pillow and
Simoncelli, 2003), in which a pre-filtering of the white noise input
results in a stable limit STA.

In this article, we focus on the problem of estimating the optimal
conductance patterns \corr{required for spike initiation, based
solely on the} analysis of V$_m$ activity.  We consider neurons
subject to conductance-based synaptic noise at both excitatory and
inhibitory synapses.  By discretizing the time axis, it is possible
to obtain the probability distribution of conductance time courses
that are compatible with the observed voltage STA.  Due to the
symmetry properties of the probability distribution, the STA time
course of excitatory and inhibitory conductances can then be
extracted by choosing the one with maximum likelihood.  We test this
method in numerical simulations of the IF model, as well as in real
cortical neurons using the dynamic-clamp technique, \corr{by
comparing the estimated STA with the real STA deduced from the
injected conductances.}

% Some of these results were presented in conference
% abstracts~\cite{conf-abstracts}. 
% WHICH OTHER ABSTRACTS SHOULD WE MENTION HERE?

% \preprint{\clearpage}

%---------------------- MATERIAL AND METHODS ----------------------
% Describe techniques, cell/animal models used, and lists of reagents,
% chemicals, and equipment, as well as the names of manufacturers and
% suppliers, so that your study can be most easily replicated by
% others. Also in this section, describe the statistical methods that
% were used to evaluate the data. For all investigations involving
% humans or animals, a statement of protocol approval from an IRB or
% IACUC, or an equivalent statement, must be included (see Use of
% Humans and/or Animals in Experiments).

\subsection*{Material and Methods}

\subsubsection*{\it Models}

We considered neurons driven by synaptic noise described by two
independent sources of colored conductance noise (point-conductance
model (Destexhe et al., 2001)).  The membrane equation of this system
is given by:

\begin{eqnarray}
    \label{eq:vm}
    C \frac{dV(t)}{dt} &=& -g_L \big( V(t)-V_L \big) -g_e(t) \big( V(t)-V_e \big) -g_i(t) \big( V(t)-V_i \big) + I_{DC},  \\
    \label{eq:gs}
    \frac{dg_s(t)}{dt} &=& -\frac{1}{\tau_s} \big( g_s(t)-g_{s 0}\big) + \sqrt{\frac{2 \sigma_s^2}{\tau_s}} \xi_s(t), 
\end{eqnarray}
Here, $g_L$, $g_e(t)$ and $g_i(t)$ are the conductances of leak,
excitatory and inhibitory currents, $V_L$, $V_e$, $V_i$ are their
respective reversal potentials, $C$ is the capacitance and $I_{DC}$ a
constant current. The subscript $s$ in Eq.~(\ref{eq:gs}) can take the
values $e, i$, which in turn indicate the respective excitatory or
inhibitory channel. We use $g_{s 0}$ and $\sigma_s$ to indicate the
mean and standard deviation (SD) of the conductance distributions,
$\xi_s(t)$ are gaussian white noise processes with zero mean and unit
standard deviation. Throughout this article we use the correlation
times $\tau_e = 2.728$ ms and $\tau_i = 10.49$ ms.

This system was solved using numerical simulations of the leaky IF
model, which was adjusted to match recordings of cortical neurons in
slices (threshold $-55$ mV, refractory period $3$ ms, reset $-75$ mV). 
Simulations were done using the NEURON simulation environment (Hines
and Carnevale, 1997).  To calculate STAs, approximately 1000 spikes 
occurring during spontaneous activity were used, each
being preceded by a period of at least $100$~ms of silence to avoid 
``contamination'' of the V$_m$ STA by preceding spikes.  The same
analysis protocols (see Results) were applied to the model and to
experimental data.

\corr{In order to address the influence of a dendritic filter on the
reliability of our method, we used a two-compartment model based on
that by Pinsky and Rinzel (1994). We removed all active channels and
replaced them by an integrate-and-fire mechanism at the soma. The
geometry (L = 3.18~$\mu$m, diam = 10~$\mu$m) as well as the
parameters not related to the spiking mechanism ($g_L =
10^{-4}$~S/cm$^2$, $V_L = -60$~mV, capacitance $c_m = 3~\mu$F/cm$^2$,
axial resistance $R_a = 5.87 \times 10^{5}~\Omega$cm) are identical
for the two compartments. In addition, we chose a threshold for
spiking V$_t = -55$~mV at the soma. The parameters for leak
conductance and capacitance needed for the estimation of the STAs of
synaptic conductances from the V$_m$, $g^{so}_L$ and $C^{so}$, were
obtained by current pulse injection into the soma at the resting
state. The superscript $so$ indicates that these are effective values
at the level of the soma.  The values used were $g^{so}_L$ = 0.198~nS
and $C^{so}$ = 5.86~pF. The parameters describing the distributions
of synaptic conductances were chosen in a way such that the mean
inhibitory conductance was four times that of excitation, and the
latter was comparable to the leak conductance ($g_{e0}$ = 0.15~nS,
$g_{i0}$ = 0.6~nS). Standard deviations were assumed to be one third
of the respective means ($\sigma_e$ = 0.05~nS, $\sigma_i$ = 0.2~nS).}

\subsubsection*{\it In vitro experiments}

{\it In vitro} experiments were performed on 0.4 mm thick coronal or
sagittal slices from the lateral portions of guinea-pig occipital
cortex.  Guinea-pigs, 4-12 weeks old (CPA, Olivet, France), were
anesthetized with sodium pentobarbital (30 mg/kg).  The slices were
maintained in an interface style recording chamber at 33-35$\degC$. 
Slices were prepared on a DSK microslicer (Ted Pella Inc., Redding,
CA) in a slice solution in which the NaCl was replaced with sucrose
while maintaining an osmolarity of 307 mOsm.  During recording, the
slices were incubated in slice solution containing (in mM): NaCl,
124; KCl, 2.5; MgSO$_4$, 1.2; NaHPO$_4$, 1.25; CaCl$_2$, 2;
NaHCO$_3$, 26; dextrose, 10, and aerated with 95\% O$_2$, 5\% CO$_2$
to a final pH of 7.4.  Intracellular recordings following two hours
of recovery were performed in deep layers (layer IV, V and VI)
\corr{in} electrophysiologically identified regular spiking and
intrinsically bursting cells.  Electrodes for intracellular
recordings were made on a Sutter Instruments P-87 micropipette puller
from medium-walled glass (WPI, 1BF100) and beveled on a Sutter
Instruments beveler (BV-10M).  Micropipettes were filled with 1.2 to
2 M potassium acetate and had resistances of 80-100 M$\Omega$ after
beveling.  

The dynamic-clamp technique (Robinson et al., 1993; Sharp et al.,
1993) was used to inject computer-generated conductances in real
neurons.  Dynamic-clamp experiments were run using the hybrid
RT-NEURON environment (developed by G.\ Le Masson, Universit\'e de
Bordeaux), which is a modified version of NEURON (Hines and
Carnevale, 1997) running under the Windows 2000 operating system
(Microsoft Corp.).  NEURON was augmented with the capacity of
simulating neuronal models in real time, synchronized with the
intracellular recording.  To achieve \corr{real-time} simulations as
well as data transfer to the PC for further analysis, we used a PCI
DSP board (Innovative Integration, Simi Valley, USA) with 4
analog/digital (inputs) and 4 digital/analog (outputs) 16 bits
converters.  The DSP board \corr{constrains} calculations of the
models and data \corr{transfers} to be made with a high priority
level by the PC processor.  The DSP board allows input (for instance
the membrane potential of the real cell incorporated in the equations
of the models) and output signals (the synaptic current to be
injected into the cell) to be processed at regular intervals (time
resolution = 0.1~ms).  A custom interface was used to connect the
digital and analog inputs/outputs signals of the DSP board with the
intracellular amplifier (Axoclamp 2B, Axon Instruments) and the data
acquisition systems (PC-based acquisition software ELPHY, developed
by G.\ Sadoc, CNRS Gif-sur-Yvette, ANVAR and Biologic).  The
dynamic-clamp protocol was used to insert the fluctuating
conductances underlying synaptic noise in cortical neurons using the
point-conductance model, similar to a previous study (Destexhe et
al., 2001).  According to Eq.~(\ref{eq:vm}) above, the injected
current is determined from the fluctuating conductances $g_e(t)$ and
$g_i(t)$ as well as from the difference of the membrane voltage
\corr{from} the respective reversal potentials, $I_{DynClamp} = -g_e
\big( V - V_e \big) -g_i \big( V - V_i \big)$.

All research procedures concerning the experimental animals and their
care adhered to the American Physiological Society's Guiding
Principles in the Care and Use of Animals, to the European Council
Directive 86/609/EEC and to European Treaties series no.\ 123, and
was also approved by the local ethics committee ``Ile-de-France Sud''
(certificate no.\ 05-003).

% \preprint{\clearpage}

%---------------------- RESULTS ------------------------------------
% Provide the experimental data and results as well as the particular
% statistical significance of the data.

\subsection*{Results}

We first \corr{explain} the method \corr{for extracting} STAs from
V$_m$ activity, then we present tests of this method using numerical
simulations and intracellular recordings in dynamic-clamp.  

\subsubsection*{\it Method to extract conductance STA}

The procedure we follow here to estimate STA of conductances from
V$_m$ activity is based on a discretization of the time axis.  With
this approach, a probability distribution can be constructed whose
maximum gives the most likely conductance path compatible with the
STA of the V$_m$.  This maximum is determined by a system of linear
equations which is solvable if the means and variances of
conductances are known \corr{(for a method to estimate conductance
mean and variance, see Rudolph et al., 2004).}

We start from the voltage STA, which is an average over an ensemble
of event-triggered voltage traces.  Its relation to the conductance
STAs is determined by the ensemble average of Eqs.~(\ref{eq:vm}) and
(\ref{eq:gs}).  In general, there is a strong correlation (or
anti-correlation) between $V(t)$ and $g_s(t)$ in time.  However, it
is safe to assume that there is no such correlation across the
ensemble, since the noise processes $\xi_s(t)$ corresponding to each
realization are uncorrelated. Also, the ensemble average \corr{is
commutative} with the time derivative.  Thus, we can rewrite
Eqs.~(\ref{eq:vm}) and (\ref{eq:gs}) to obtain

\begin{eqnarray}
    \label{eq:vmav}
    \frac{d \langle V(t) \rangle_x}{dt} &=& -\frac{1}{\tau_L} \big( \langle V(t) \rangle_x -V_L \big) \\
    && -\frac{\langle g_e(t) \rangle_x}{C} \big( \langle V(t) \rangle_x -V_e \big) -\frac{\langle g_i(t) \rangle_x}{C} \big( \langle V(t) \rangle_x -V_i \big) + \frac{I_{DC}}{C}, \nonumber \\
    \label{eq:gsav}
    \frac{d \langle g_s(t) \rangle_x}{dt} &=& -\frac{1}{\tau_s} \big( \langle g_s(t) \rangle_x -g_{s 0} \big) + \sqrt{\frac{2 \sigma_s^2}{\tau_s}} \langle \xi_s(t) \rangle_x, 
\end{eqnarray}
where $\tau_L = C / g_L$ and $\langle . \rangle_x$ denotes the
ensemble average.  In other words, the time evolution
Eqs~(\ref{eq:vm}) and (\ref{eq:gs}) also hold in terms of ensemble
averages. In the following, we drop the bracket notation for
legibility, but assume we are dealing with ensemble averaged
quantities unless otherwise stated.  

We discretize Eq.~(\ref{eq:vmav}) in time with a step-size $\Delta t$
and solve for $g_i^k$,
\begin{equation}
    \label{eq:gi}
    g_i^k = -\frac{C}{V^k-V_i} \left\{ \frac{V^k-V_L}{\tau_L} + \frac{g_e^k (V^k - V_e)}{C} + \frac{V^{k+1} - V^k}{\Delta t} -\frac{I_{DC}}{C} \right\}.
\end{equation}
Since the series $V^k$ for the voltage STA is known, $g_i^k$ has
become a function of $g_e^k$. In the same way, we solve
Eq.~(\ref{eq:gsav}) for $\xi_s^k$, which have become gaussian
distributed random numbers,
\begin{equation}
    \label{eq:xi}
    \xi_s^k = \frac{1}{\sigma_s} \sqrt{\frac{\tau_s}{2 \Delta t}} \left( g_s^{k+1} - g_s^k \Big( 1 - \frac{\Delta t}{\tau_s} \Big) - \frac{\Delta t}{\tau_s} g_{s 0} \right).
\end{equation}
There is a continuum of combinations $\{ g_e^{k+1}, g_i^{k+1}\}$ that
can advance the membrane potential from $V^{k+1}$ to $V^{k+2}$, each
pair occurring with a probability
\begin{eqnarray}
    p^k &:=& p(g_e^{k+1}, g_i^{k+1} | g_e^k, g_i^k) = \frac{1}{2 \pi} e^{-\frac{1}{2} (\xi_e^{k 2} + \xi_i^{k 2})} = \frac{1}{2 \pi} e^{-\frac{1}{4 \Delta t} X^k}, \\
    X^k &=& \frac{\tau_e}{\sigma_e^2} \left( g_e^{k+1} - g_e^k \Big( 1 - \frac{\Delta t}{\tau_e} \Big) - \frac{\Delta t}{\tau_e} g_{e 0} \right)^2 \\ 
    && + \frac{\tau_i}{\sigma_i^2} \left( g_i^{k+1} - g_i^k \Big( 1 - \frac{\Delta t}{\tau_i} \Big) - \frac{\Delta t}{\tau_i} g_{i 0} \right)^2, \nonumber
\end{eqnarray}
where we have used Eq.~(\ref{eq:xi}). Note that because of
Eq.~(\ref{eq:gi}),  $g_e^k$ and $g_i^k$ are not independent and $p^k$
is, thus, a unidimensional distribution only. Given initial
conductances $g_s^0$, we can now write down the probability $p$ for
certain series of conductances $\{g_s^j\}_{j = 0, \ldots, n}$ to
occur that reproduce a given voltage trace $\{ V^l \}_{l = 1, \ldots,
n+1}$:
\begin{equation}
    p = \prod_{k = 0}^{n-1} p^k.
\end{equation}
Due to the symmetry of the distribution $p$, the average paths of the
conductances coincide with the most likely ones, so the cumbersome
task of solving nested gaussian integrals can be circumvented.
Instead, in order to determine the conductance series with extremal
likelihood, we solve the n-dimensional system of linear equations
\begin{equation}
    \label{eq:sys}
    \left\{ \frac{\partial X}{\partial g_e^k} = 0\right\}_{k = 1, \ldots, n},
\end{equation}
where $X = \sum_{k = 0}^{n-1} X^k$, for the vector $\{ g_e^k \}$.
This is equivalent to solving $\{\frac{\partial p}{\partial g_e^k} =
0\}_{k = 1, \ldots, n}$ and involves the numerical inversion of
\corr{an} $n \times n$-matrix. Since the system of equations is
linear, if there is a solution for $\{ g_e^k \}$, plausibility
arguments suggest that it is the most likely (rather than the least
likely) excitatory conductance time course. The series $\{ g_i^k \}$
is then obtained from Eq.~(\ref{eq:gi}).  

\subsubsection*{\it Test of the accuracy of the method using
numerical simulations}

To test this method, we first considered numerical simulations of the
IF model in four different situations. We distinguished
high-conductance states, where the total conductance is dominated by
inhibition, from low-conductance states, where both synaptic
conductances are of comparable magnitude. We also varied the standard
deviations of the conductances such that for both high- and
low-conductance states we have the cases $\sigma_i > \sigma_e$ as
well as $\sigma_e > \sigma_i$. The results are summarized in
Fig.~{\ref{fig:if}}, where the STA traces of excitatory and
inhibitory conductances recorded from simulations are compared to the
most likely (equivalent to the average) conductance traces obtained
from solving Eq.~(\ref{eq:sys}). In general, the plots demonstrate a
very good agreement.

To quantify our results, we investigated the effect of \corr{the}
statistics as well as of the broadness of the conductance
distributions on the quality of the estimation.  The latter is
crucial, because the derivation of the most likely conductance time
course allows for negative conductances, whereas in the simulations
negative conductances lead to numerical instabilities, and
conductances are bound to positive values. We thus expect an
increasing error with increasing ratio SD/mean of the conductance
distributions.  We estimated the root-mean-square (RMS) of the
difference between the recorded and the estimated conductance STAs. 
The results, summarized in Fig.~\ref{fig:stat}, are as expected.
Increasing the number of spikes enhances the match between theory and
simulation (Fig.~\ref{fig:stat}A shows the RMS \corr{deviation} for
excitation, \corr{\ref{fig:stat}B} for inhibition) up to the point
where the effect of negative conductances becomes dominant. In this
example, where the ratio SD/mean was fixed at $0.1$, the RMS
\corr{deviation} enters a plateau at about $7000$ spikes.  The
plateau \corr{values} can also be recovered from the neighboring
plots (\corr{i.e., the RMS deviations at SD/mean $=0.1$ in}
Fig.~\ref{fig:stat}\corr{C} and D \corr{correspond to the plateau
values in A and B}). On the other hand, a broadening of the
conductance distribution yields a higher deviation between simulation
and estimation. However, at SD/mean $= 0.5$, the RMS \corr{deviation}
is still as low as $\sim 2\%$ of the mean conductance for excitation
and $\sim 4\%$ for inhibition.

\corr{To assess the effect of dendritic filtering on the reliability
of the method, we used a two-compartment model based on that of
Pinsky and Rinzel (1994), from which we removed all active channels
and replaced them by an integrate-and-fire mechanism at the soma.  We
repeatedly injected the same 100~s sample of fluctuating excitatory
and inhibitory conductances in the dendritic compartment and
performed two different recording protocols at the soma
(Fig.~\ref{fig:atten}A).  First, we recorded in current-clamp in
order to obtain the V$_m$ time course as well as the spike times.  In
this case, the leak conductance $g^{so}_L$ and the capacitance
$C^{so}$ were obtained from current pulse injection at rest. Second,
we simulated an ``ideal'' voltage-clamp (no series resistance) at the
soma using two different holding potentials (we chose the reversal
potentials of excitation and inhibition, respectively). Then, from
the currents $I_{V_e}$ and $I_{V_i}$, one can calculate the
conductance time courses as
\begin{equation}
    \label{eq:vclamp}
    g^{so}_{e,i}(t) = \frac{I_{V_{i,e}}(t)-g_L (V_{i,e}-V_L)}{V_{i,e}-V_{e,i}},
\end{equation}
where the superscript $so$ indicates that these are the conductances
seen at the soma (in the following referred to as somatic
conductances).  From these, we determined the parameters $g^{so}_{e
0}$, $g^{so}_{i 0}$, $\sigma^{so}_e$ and $\sigma^{so}_i$, the
conductance means and standard deviations. In contrast to $g_e(t)$
and $g_i(t)$, the distributions of $g^{so}_e(t)$ and $g^{so}_i(t)$
are not Gaussian (not shown), and have lower means and variances.  We
compared the STA of the injected (dendritic) conductance, the STA
obtained from the somatic V$_m$ using our method and the STA obtained
using a somatic ``ideal'' voltage-clamp (see
Fig.~\ref{fig:atten}B-D), which demonstrated the following points:
(1) as expected, due to dendritic attenuation, all somatic estimates
were attenuated compared to the actual conductances injected in
dendrites (compare light and dark gray curves, soma, with black
curve, dendrite, in Fig.~\ref{fig:atten}B-D); (2) the estimate
obtained by applying the present method to the somatic V$_m$ (dark
gray curves in Fig.~\ref{fig:atten}B-D) was very similar to that
obtained using an ``ideal'' voltage-clamp at the soma (light gray
curves).  The difference close to the spike may be due to the
non-Gaussian shape of the somatic conductance distributions, whose
tails then become important;  (3) despite attenuation, the
qualitative shape of the conductance STA was preserved.  We conclude
that the STA estimate from V$_m$ activity captures rather well the
conductances as seen by the spiking mechanism.}

\subsubsection*{\it Test of the method in real neurons}

We also tested the method on voltage STAs obtained from dynamic-clamp
recordings of guinea-pig cortical neurons in slices.  In real
neurons, a problem is the strong influence of spike-related
voltage-dependent (presumably sodium) conductances on the voltage
time course. Since we maximize the {\it global} probability of
$g_e(t)$ and $g_i(t)$, the voltage in the vicinity of the spike has
an influence on the estimated conductances at all times. As a
consequence, without removing the effect of sodium, the estimation
fails (see Fig.~\ref{fig:sodium}).  Fortunately, it is rather simple
to correct for this effect by excluding the last 1--2 ms before the
spike from the analysis. The corrected comparison between the
recorded and the estimated conductance traces is shown in
Fig.~\ref{fig:dc}.

Finally, to check for the applicability of this method to {\it in
vivo} recordings, we assessed the sensitivity of the estimates with
respect to the different parameters \corr{by varying} the values
describing passive properties and synaptic activity.  We assume that
the total conductance can be constrained by input resistance
measurements, and that time constants of the synaptic currents can be
estimated by power spectral analyses (Destexhe and Rudolph, 2004). 
This leaves $g_L$, $C$, $g_{e 0}$, $\sigma_e$ and $\sigma_i$ as the
main parameters.  The impact of these parameters on STA conductance
estimates is shown in Fig.~\ref{fig:deviations}.  Varying these
parameters within $\pm 50 \%$ of their nominal value led to various
degrees of error in the STA estimates.  The dominant effect of a
\corr{variation} in the mean conductances is a shift in the estimated
STAs, whereas a \corr{variation} in the SDs \corr{changes} the
curvature just before the spike.  

To address this point further, we fitted the estimated conductance
STAs with an exponential function: 

\begin{equation}
    f_s(t) = G_s (1 + K_s e^{\frac{t-t_0}{T_s}}), \nonumber
\end{equation}
where $s$ again takes the values $e, i$ for excitation and
inhibition, respectively. $t_0$ is chosen to be the time \corr{at
which} the analysis stops. Fig.~\ref{fig:deviations_details} gives an
overview of the dependence of the fitting parameters $G_e$, $G_i$,
$T_e$ and $T_i$ on the relative change of $g_L$, $g_{e 0}$,
$\sigma_e$, $\sigma_i$ and $C$.  For example, a \corr{variation of}
$g_{e 0}$ has a strong effect on $G_e$ and $G_i$, but \corr{affects}
to a \corr{lesser} extent $T_e$ and $T_i$, while the opposite was
seen \corr{when varying} $\sigma_e$ and $\sigma_i$.

% \preprint{\clearpage}

%---------------------- DISCUSSION --------------------------------
% Explain your interpretation of the data, especially compared with
% previously published material cited in the References.

\subsection*{Discussion}

Understanding the transfer function of a neuron from synaptic input
to spike output would ideally require the simultaneous monitoring of
both the synaptic conductances and the cell's firing. Current methods
for extracting synaptic conductances rely on intracellular recordings
performed at different holding potentials (in voltage-clamp) or
different current levels (in current-clamp; e.g.\ Borg-Graham et al. 
1998) and, as a consequence, they do not allow \corr{the
establishment of} a direct correspondence between synaptic
conductances and spikes.  Although these methods have been very
useful, for example \corr{in establishing} the synaptic structure of
sensory receptive fields in a variety of systems (Monier et al.,
2003; Wehr and Zador, 2003; Wilent and Contreras, 2005), they do not
distinguish between trials that effectively produce spikes at a given
latency and those that do not.  

Here, we have presented a method to extract the average excitatory
and inhibitory conductance patterns directly related to \corr{spike
initiation}.  As illustrated in Fig.~\ref{fig:sc}, this method can
extract spike-related conductances based \corr{solely} on the
knowledge of V$_m$ activity.  First, the STA of the V$_m$ is computed
from the intracellular recordings.  Next, by discretizing the time
axis, one estimates the ``most likely'' conductance time courses that
are compatible with the observed STA of V$_m$.  Due to the symmetry
of \corr{their} distribution, the average conductance time courses
coincide with the most likely ones, so integration over the entire
stimulus space (whose dimension depends on the STA interval as well
as \corr{on} the temporal resolution) can be replaced by a
differentiation and subsequent solution of a system of linear
equations.  Solving this system gives \corr{an estimate of the
average} conductance time courses.  We demonstrated that this
estimation gives reasonably accurate estimates for the leaky IF
model, as well as in real neurons under dynamic-clamp.

Like any other method, this method suffers from several sources of
error.  Errors can result from nonlinearities in the I-V curve of the
neuron, e.g.\ \corr{those} due to voltage-dependent conductances.  In
agreement with this, we have shown that the subthreshold activation
of spike-generating currents close to threshold can lead to severe
\corr{misestimations of the} conductances (Fig.~\ref{fig:sodium}). 
This problem can be circumvented by excluding a short period
(1--2~ms) preceding the spike.  To avoid contamination by
voltage-dependent currents, this method should be complemented by a
check for I-V curve linearity in the range of V$_m$ considered. 
\corr{Note that a linear I-V curve does not guarantee the absence of
voltage-dependent conductances.  For example, if the mean interspike
interval of the cell becomes too short, spike-related potassium
currents might be present during a substantial fraction of the STA
interval and could affect the estimation. This might diminish the
applicability of the method to neurons spiking at high frequency, in
particular to fast-spiking interneurons. Also, strong subthreshold
dendritic conductances that are very remote from the soma could
influence the STA estimate without being visible in the I-V curve. On
the other hand, in cases where it is possible to parameterize these
nonlinearities, they can be included in Eq.~\ref{eq:gi}. It should
thus be possible to extend the method in order to apply it to more
complex models, for example the exponential integrate-and-fire model
(Fourcaud-Trocme et al., 2003).  Another possible extension would be
to include voltage-dependent terms such as N-methyl-D-aspartate
(NMDA) receptor-mediated synaptic currents, although such currents
probably have a limited contribution at the range of V$_m$ considered
here (below -50~mV).}

Another source of error may arise from ``negative conductances''. 
The present model of synaptic noise assumes that conductances are
Gaussian-distributed, but if the standard deviation becomes
comparable to the mean value of the conductances, the Gaussian
distribution will \corr{include negative conductances, which are
unrealistic}.  This is an important limitation of representing
synaptic conductances by Gaussian-distributed noise (``diffusion
approximation'').  However, this type of approximation seems to apply
well to cortical neurons {\it in vivo}, which receive a large number
of inputs (Destexhe et al., 2001).  {\it In vivo} measurements so far
indicate that the standard deviation is much smaller than the mean
for both excitatory and inhibitory conductances (Haider et al., 2006;
Rudolph et al., 2005), which also indicates that the diffusion
approximation is valid in this case.  Such a check for consistency is
a prerequisite for applying the present method.

Previous work related to the question of spike-triggered stimuli was
mainly focused on white noise current inputs, and showed that no
stable finite input average exists in the limit $dt \rightarrow 0$
(Paninski 2006a).  Other contributions shed light on the question of
membrane voltage STAs, for the leaky IF neuron as well as for
biophysically more plausible models.  However, so far no procedure
was proposed to solve this problem of reverse correlation for
conductance noise inputs.  The method we propose here attempts to
fill this gap, and directly provides a procedure that can be applied
to real neurons.  To this end, the present method must be
complemented by measurements of the mean and standard deviation of
excitatory and inhibitory conductances.  Such measurements can be
obtained either by voltage-clamp (Haider et al., 2006), or by
current-clamp, as recently proposed (Rudolph et al., 2004, 2005). 
Combining the latter method with the present method, it should now be
possible to directly extract conductance patterns from V$_m$
recordings {\it in vivo}, and thus obtain estimates of the
conductance variations related to spikes during natural network
states.

% \preprint{\clearpage}

%---------------------- ACKNOWLEDGMENTS --------------------------
% List the people indirectly involved with the research to whom you may
% wish to give thanks. Also, current addresses of authors (if they
% differ from those in the affiliation line) should be included here.
% Do not include "promissory notes." APS Journal policy is against
% inclusion of implicit or explicit promises that future work will be
% published.  Do not include dedications. Dedications of articles are 
% not permitted.
% List the grants, fellowships, and donations that funded (partially or
% completely) the research. However, industry-sponsored grants should
% be listed under Disclosures.

\subsection*{Acknowledgments}

We thank Romain Brette for discussions and Andrew Davison for
comments on the manuscript.  Research supported by the CNRS, the ANR
(HR-CORTEX project), the Human Frontier Science Program, and the
European Community (FACETS project IST 015879).

%---------------------- REFERENCES --------------------------------

\preprint{\clearpage}

\subsection*{References}

%\begin{enumerate}
\begin{description}

\item Ag\"uera y Arcas B. and Fairhall A. L.  What causes a
neuron to spike?   {\it Neural Computation} {\bf 15}: 1789-1807, 2003.

\item Badel L., Gerstner W. and Richardson M.J.E.  Dependence of
the spike-triggered average voltage on membrane response properties. 
{\it Neurocomputing} {\bf 69}, 1062-1065, 2006.

\item Borg-Graham L., Monier C. and Fr\'egnac Y.    Visual input
evokes transient and strong shunting inhibition in visual cortical
neurons.  {\it Nature} {\bf 393}: 369-373, 1998.

\item de Boer F. and Kuypers P.  Triggered Correlation.  
{\it IEEE Trans. Biomed. Eng.} {\bf 15}: 169-197, 1968.

\item Destexhe A. and Rudolph M.  Extracting information from the
power spectrum of synaptic noise.  {\it J. Computational Neurosci.}
{\bf 17}: 327-345, 2004.

\item Destexhe A., Rudolph M. and Par\'e D.  The high-conductance
state of neocortical neurons {\it in vivo}.  {\it Nature Reviews
Neurosci.} {\bf 4}: 739-751, 2003.

\item Destexhe A., Rudolph M., Fellous J.-M. and Sejnowski T. J.   
Fluctuating synaptic conductances recreate in-vivo-like
activity in neocortical neurons.  {\it Neuroscience} {\bf 107}:
13-24, 2001.

\item \corr{Fourcaud-Trocme N., Hansel D., van Vreeswijk C., Brunel
N. How spike generation mechanisms determine the neuronal response to
fluctuating inputs. {\it J.  Neurosci.} {\bf 23}(37): 11628-11640,
2003.}

\item Haider B., Duque A., Hasenstaub A.R. and McCormick D.A.  
Network activity in vivo is generated through a dynamic balance of
excitation and inhibition. {\it J.  Neurosci.} {\bf 26}: 4535-4545,
2006.

% \item Hasenstaub A., Shu Y., Haider B., Kraushaar U., Duque A. and
% McCormick D.A.  Inhibitory postsynaptic potentials carry synchronized
% frequency information in active cortical networks. {\it Neuron} {\bf
% 47}: 423-435, 2005.

\item Hines M. L. and Carnevale N. T. (1997) The NEURON simulation
environment. {\it Neural Computation} {\bf 9}: 1179-1209, 1997

\item Monier C., Chavane F., Baudot P., Borg-Graham L. and Fr\'egnac
Y.  Orientation and direction selectivity of synaptic inputs in
visual cortical neurons: a diversity of combinations produces spike
tuning. {\it Neuron} {\bf 37}(4): 663-680, 2003.

\item Paninski L.  The spike-triggered average of the
integrate-and-fire cell driven by gaussian white noise.   {\it Neural
Computation} {\bf 18}: 2592-2616, 2006a.

\item Paninski L.  The most likely voltage path and large deviations
approximations for integrate-and-fire neurons.  {\it J.  Comput. 
Neurosci.} {\bf 21}: 71-87, 2006b.

\item Paninski L. , Pillow J. and Simoncelli E.   Maximum likelihood
estimation of a stochastic integrate-and-fire neural model.   {\it
Neural Computation} {\bf 16}: 2533-2561, 2004.

\item Pillow J. and Simoncelli E.   Biases in white noise
analysis due to non-Poisson spike generation.   {\it Neurocomputing}
{\bf 52}: 109-155, 2003.

\item \corr{Pinsky P. F. and Rinzel J.   Intrinsic and network
rhythmogenesis in a reduced Traub model for CA3 neurons. {\it J. 
Comput. Neurosci.} {\bf1}(1-2): 39-60, 1994.}

\item Robinson H. P. and Kawai N.  Injection of digitally synthesized
synaptic conductance transients to measure the integrative properties
of neurons. {\it J. Neurosci. Methods} {\bf 49}(3): 157-165, 1993.

\item  Rudolph M., Pelletier J.-G., Paré D. and Destexhe A. 
Characterization of synaptic conductances and integrative properties
during electrically-induced EEG-activated states in neocortical
neurons in vivo.  {\it J. Neurophysiol.} {\bf 94}: 2805-2821, 2005.

\item Rudolph M., Piwkowska Z., Badoual M., Bal T.. and Destexhe A. 
A method to estimate synaptic conductances from membrane potential
fluctuations.   {\it J. Neurophysiol.} {\bf 91}: 2884-2896, 2004.

\item Sharp A. A., O'Neil M. B., Abbott L. F. and Marder E.  The
dynamic clamp: artificial conductances in biological neurons.  {\it
Trends Neurosci.} {\bf 16}: 389-394, 1993.

\item Wehr M. and Zador A.    Balanced inhibition underlies tuning
and sharpens spike timing in auditory cortex.  {\it Nature} {\bf 426}
(6965):442-446, 2003.  

\item Wilent W. and Contreras D.   Dynamics of excitation and
inhibition underlying stimulus selectivity in rat somatosensory
cortex.  {\it Nature Neurosci.} {\bf 8} (10):1364-1370, 2005.

\end{description}
%\end{enumerate}

\preprint{\clearpage}

%---------------------- FIGURE LEGENDS ----------------------------

% each legend is first defined as a text box (\figA, \figB, etc...)

\def\figA{Test of the STA analysis method using an IF neuron model
subject to colored conductance noise.  A. Scheme of the procedure
used.  An IF model with synaptic noise was simulated numerically
(bottom) and the procedure to estimate STA was applied to the V$_m$
activity (top).  The estimated conductance STAs from V$_m$ were then
compared to the actual conductance STA in this model.  Bottom panels:
STA analysis for different conditions, low-conductance states (B,C),
high-conductance states (D,E), with fluctuations dominated by
inhibition (B,D) or by excitation (C,E).  For each panel, the upper
graph shows the voltage STA, the middle graph the STA of excitatory
conductance, and the lower graph the STA of inhibitory conductance. 
Solid lines (grey) show the average conductance recorded from the
simulation, while the dashed line (black) represents the conductance
estimated from the V$_m$.  Parameters in B: $g_{e 0}$=6~nS, $g_{i
0}$=6~nS, $\sigma_e$=0.5~nS, $\sigma_i$=1.5~nS; C: $g_{e 0}$=6~nS,
$g_{i 0}$=6~nS, $\sigma_e$=1.5~nS, $\sigma_i$=0.5~nS; D: $g_{e
0}$=20~nS, $g_{i 0}$=60~nS, $\sigma_e$=4~nS, $\sigma_i$=12~nS; E:
$g_{e 0}$=20~nS, $g_{i 0}$=60~nS, $\sigma_e$=6~nS, $\sigma_i$=3~nS}

\def\figB{The root-mean-square (RMS) of the deviation of the
estimated from the recorded STAs. A: RMS \corr{deviation} as a
function of the number of spikes for the STA of excitatory
conductance, where the SD of the conductance distribution was $10\%$
of its mean. The RMS \corr{deviation} first decreases with the number
of spikes, but saturates at $\sim 7000$ spikes. This is due to the
effect of negative conductances, which are excluded in the simulation
(cf. \corr{C}). \corr{B: Same as A for inhibition.} \corr{C}: RMS
\corr{deviation} for excitation as a function of the ratio SD/mean of
the conductance distribution.  The higher the probability of negative
conductances, the higher the discrepancy between theory and
simulation.  However, at SD/mean $=0.5$, the mean deviation is as low
as $\sim 2\%$ of the mean conductance for excitation and $\sim 4\%$
for inhibition.  D: Same as \corr{C} for inhibition.}

\def\figC{\corr{Test of the method using dendritic conductances.  A. 
Simulation scheme: A 100~s sample of excitatory and inhibitory
(frozen) conductance noise was injected into the dendrite of a
2-compartment model (1). Then, two different recording protocols were
performed at the soma. First, the V$_m$ time course was recorded in
current-clamp (2), second, the currents corresponding to two
different holding potentials were recorded in voltage-clamp (3). From
the latter, the excitatory and inhibitory conductance time courses
were extracted using Eq.~\ref{eq:vclamp}. B. STA of total conductance
inserted at the dendrite (black), compared with the estimate obtained
in voltage-clamp (light gray) and with that obtained from somatic
V$_m$ activity using the method (dark gray).  Due to dendritic
attenuation, the total conductance values measured are lower than the
inserted ones, but the variations of conductances preceding the spike
are conserved. C. Same as B, for excitatory conductance. D. Same as
B, for inhibitory conductance. Parameters: $g_{e 0} = 0.15$ nS, $g_{i
0} = 0.6$ nS, $\sigma_e = 0.05$ nS, $\sigma_i = 0.2$ nS, $g^{so}_{e
0} = 0.113$ nS, $g^{so}_{i 0} = 0.45$ nS, $\sigma^{so}_e = 0.034$ nS,
$\sigma^{so}_i = 0.12$ nS, where the superscript $so$ denotes
quantities as seen at the soma.}}

\def\figD{The effect of the presence of additional
\corr{voltage-dependent} conductances on the estimation of the
synaptic conductances. Gray, solid lines indicate recorded
conductances; black, dotted lines indicate estimated conductances. 
In this case, the estimation fails.  The sharp rise of the voltage in
the last ms before the spike requires very fast changes in the
synaptic conductances, which introduces a  considerable error in the
analysis. Parameters used: $g_{e 0}$=32~nS, $g_{i 0}$=96~nS,
$\sigma_e$=8~nS, $\sigma_i$=24~nS.}

\begin{figure}
    \includegraphics[width=10cm]{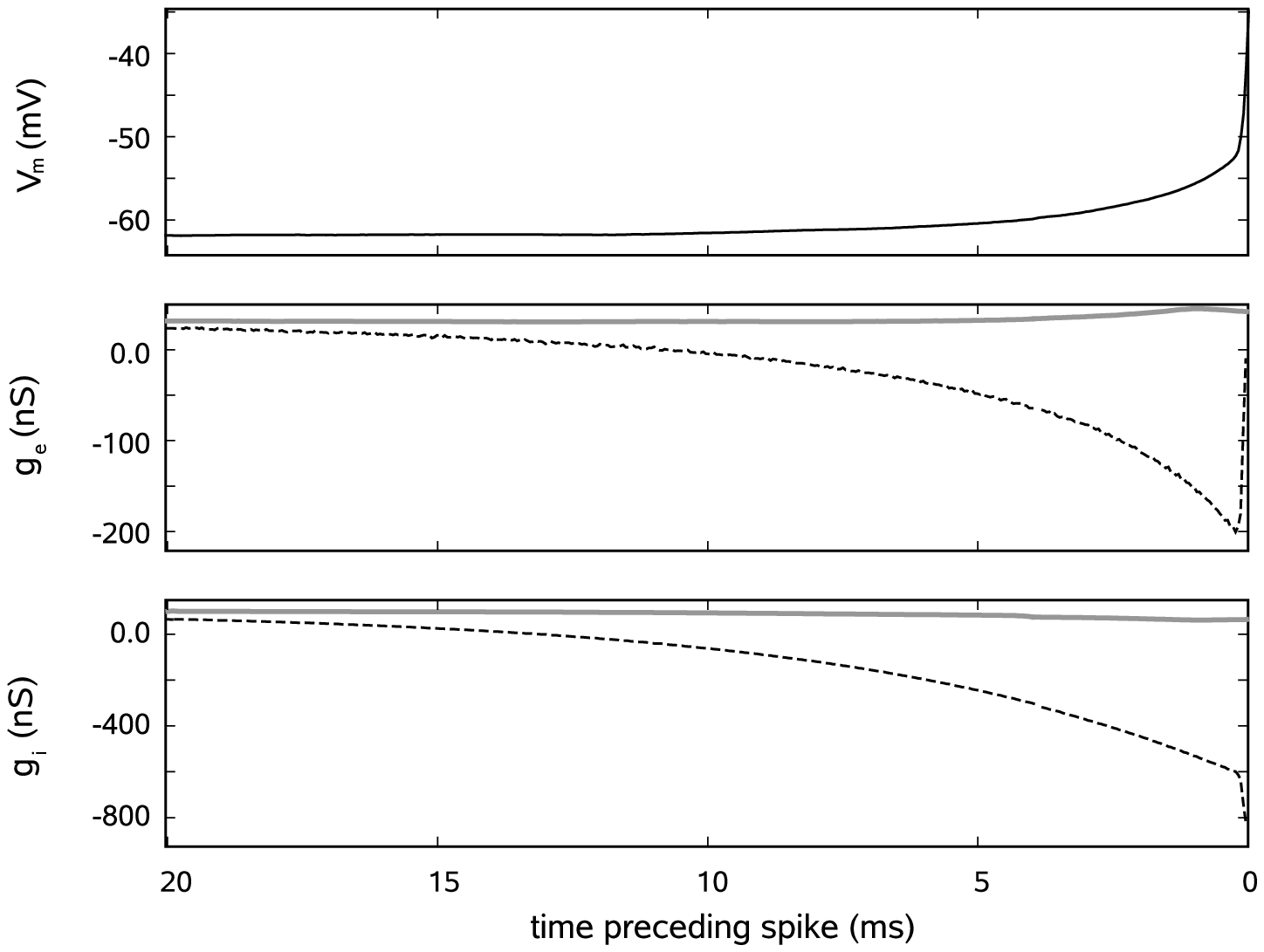}

    \caption{ \small \figD}

    \label{fig:sodium}
\end{figure}

\def\figE{Test of the method in real neurons using dynamic-clamp in
guinea-pig visual cortical slices.  A.  Scheme of the procedure. 
Computer-generated synaptic noise was injected in the recorded neuron
under dynamic-clamp (bottom).  The V$_m$ activity obtained (top) was
then used to extract the STA of conductances, which was compared to
the STA directly obtained from the injected conductances.  B. Results
of this analysis in a representative neuron.  Black lines show the
estimated STA of conductances from V$_m$ activity, grey lines show
the STA of conductances that were actually injected into the neuron. 
The analysis was made by excluding the data \corr{from the} 1.2~ms
before the spike to avoid contamination by voltage-dependent
conductances.  Parameters for conductance noise were as in
Fig.~\ref{fig:sodium}.}

\def\figF{Deviation in the estimated conductance STAs in real neurons
using dynamic-clamp due to \corr{variations} in the parameters.  The
black lines represent the \corr{conductance STA estimates using the
correct parameters}, the gray areas are bound by the estimates that
result from \corr{variation} of a single parameter (indicated on the
right) by $\pm$~50~\%. Light gray areas represent inhibition, dark
gray areas represent excitation. The total conductance (leak plus
synaptic conductances) was assumed to be fixed.  \corr{A variation}
in the mean values of the conductances evokes mostly a shift in the
estimate, while a \corr{variation} in the SDs influences the
curvature just before the spike.}

\def\figG{Detailed evaluation of the sensitivity to parameters. The
conductance STAs were fitted with an exponential function $ f_s(t) =
G_s (1 + K_s exp( (t-t_0)/T_s)$, $s = e, i$. $t_0$ is chosen to be
the time \corr{at which} the analysis stops.  Each plot shows the
estimated value of $G_e$, $G_i$, $T_e$ or $T_i$ from this experiment,
each curve represents the variation of a single parameter (see
legend).}

\def\figH{Scheme of the method to extract spike-triggered average
conductances from membrane potential activity.  Starting from an
intracellular recording (top), the spike-triggered average (STA)
membrane potential (V$_m$) is computed (leftmost panel).  From the
STA of the V$_m$, by discretizing the time axis, it is possible to
estimate the STA of conductances (bottom) by maximizing a probability
distribution (see text).  This step requires \corr{knowledge of} the
values of the average conductances and their standard deviations
($g_{e0}$, $g_{i0}$, $\sigma_e$, $\sigma_i$, respectively), which
must be extracted independently (rightmost panel).}

% This code below produces the list of figure legends; J Neurophysiol
% requires the legends to be listed before the figures.  The boxes
% are pasted together with the figure number

\preprint{\subsection*{Figure Legends}}

\begin{figure}
    \includegraphics[width=16cm]{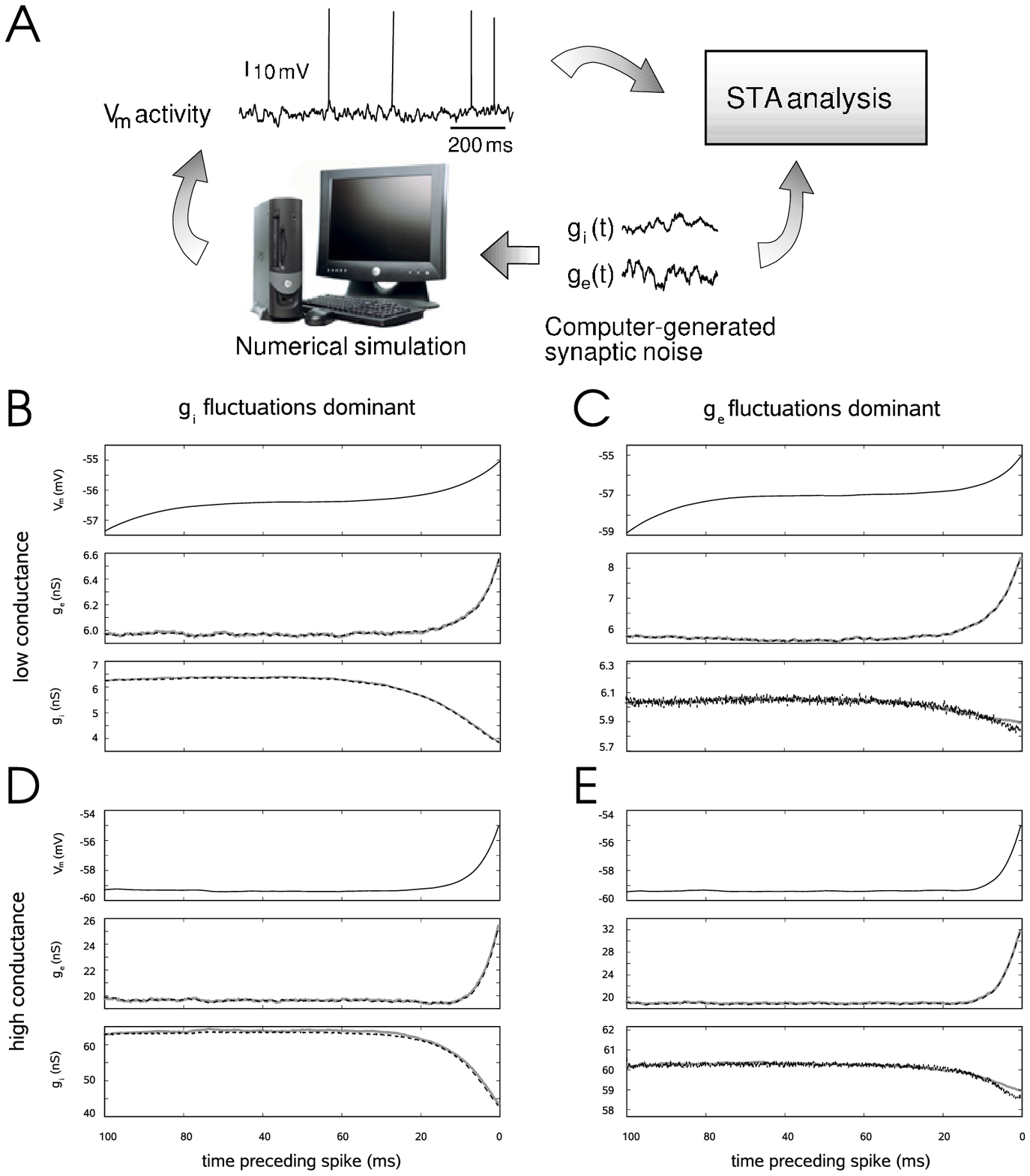}

    \caption{ \small \figA}

    \label{fig:if}
\end{figure}

\preprint{\noindent Figure~\ref{fig:if}: \figA\\ }

\begin{figure}
    \includegraphics[width=12cm]{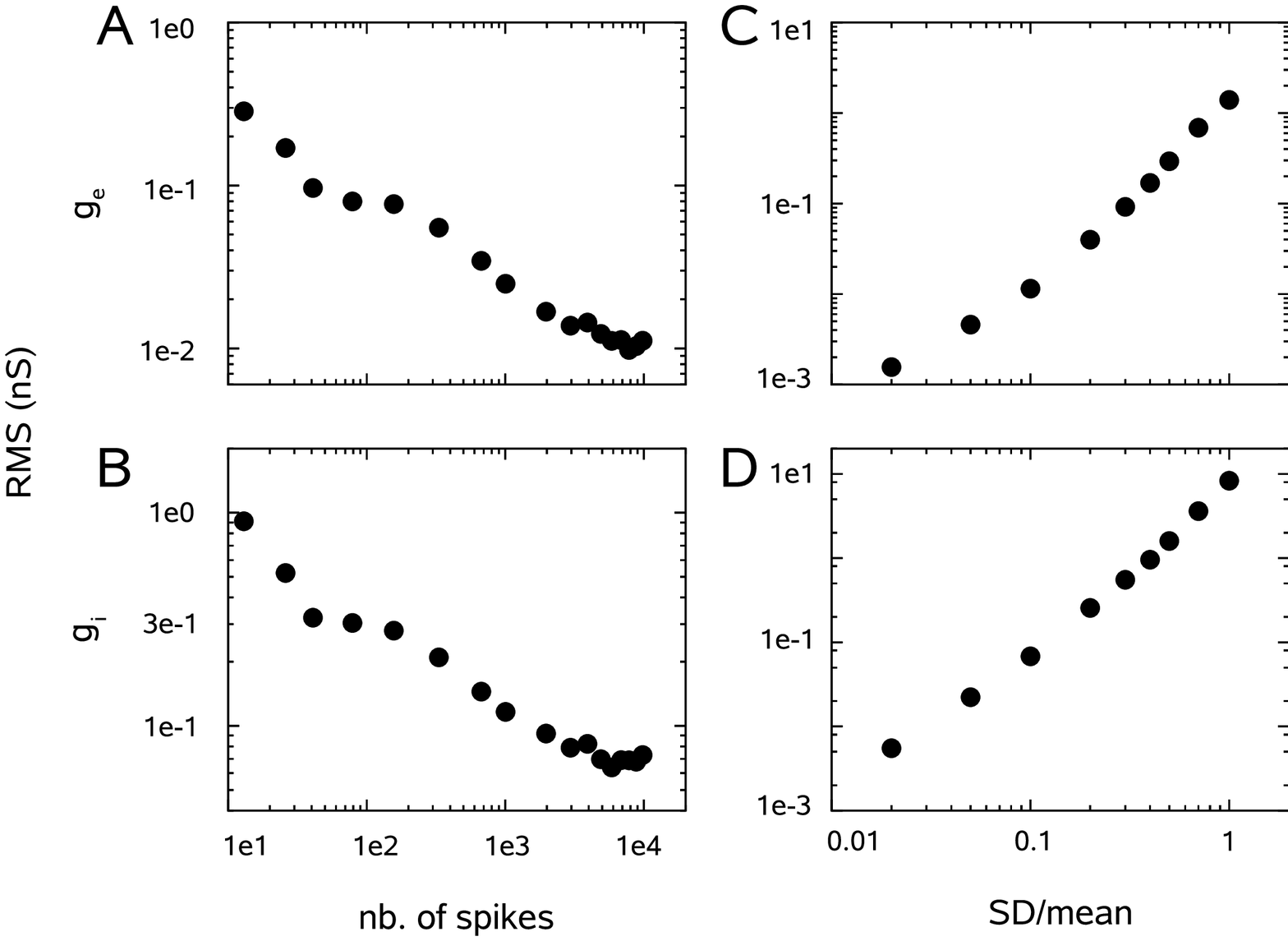}
    
    \caption{ \small \figB}
    
    \label{fig:stat}
\end{figure}

\preprint{\noindent Figure~\ref{fig:stat}: \figB\\ }

\begin{figure}
    \includegraphics[width=14cm]{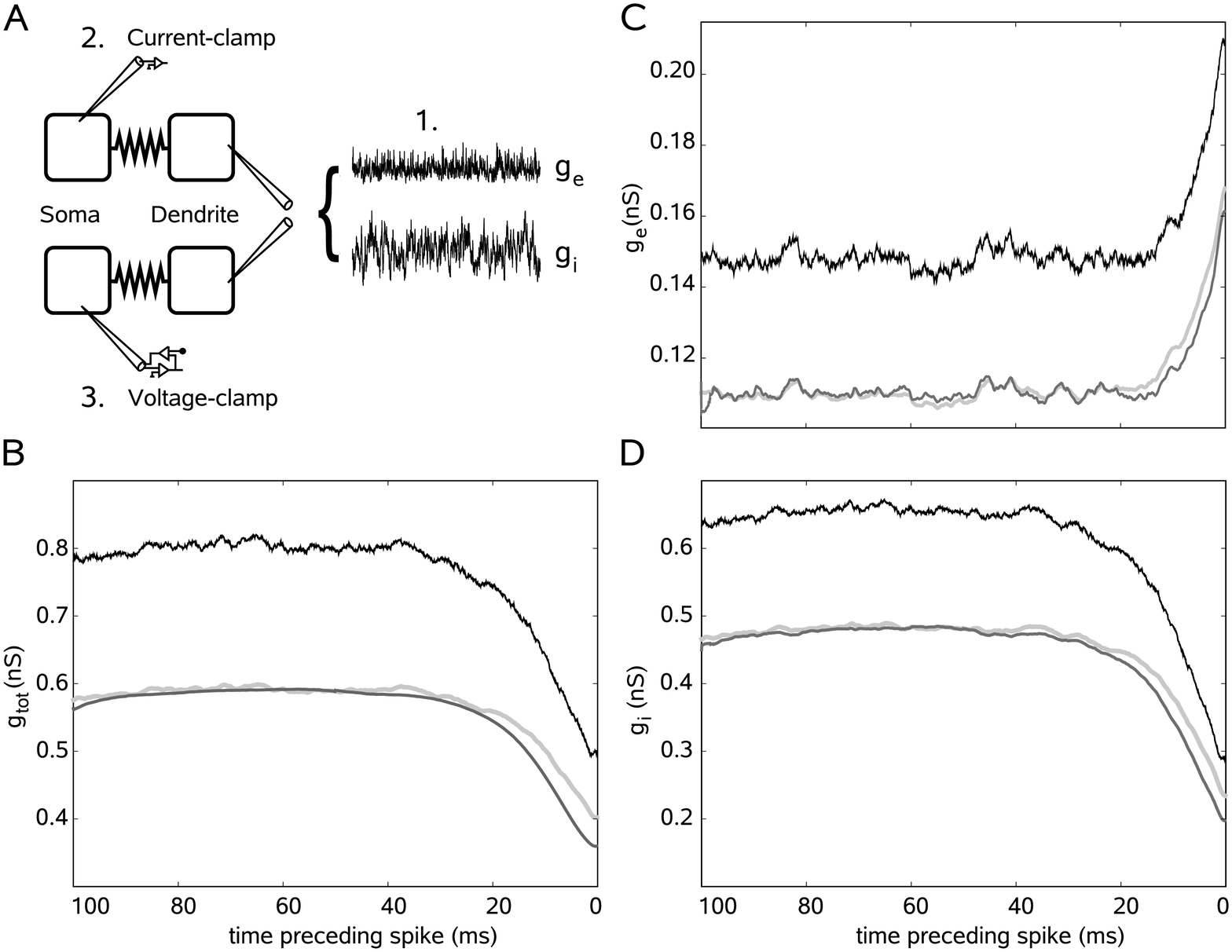}

    \caption{ \small \figC}

    \label{fig:atten}
\end{figure}

\preprint{\noindent Figure~\ref{fig:atten}: \figC\\ }

\preprint{\noindent Figure~\ref{fig:sodium}: \figD\\ }

\begin{figure}
    \includegraphics[width=14cm]{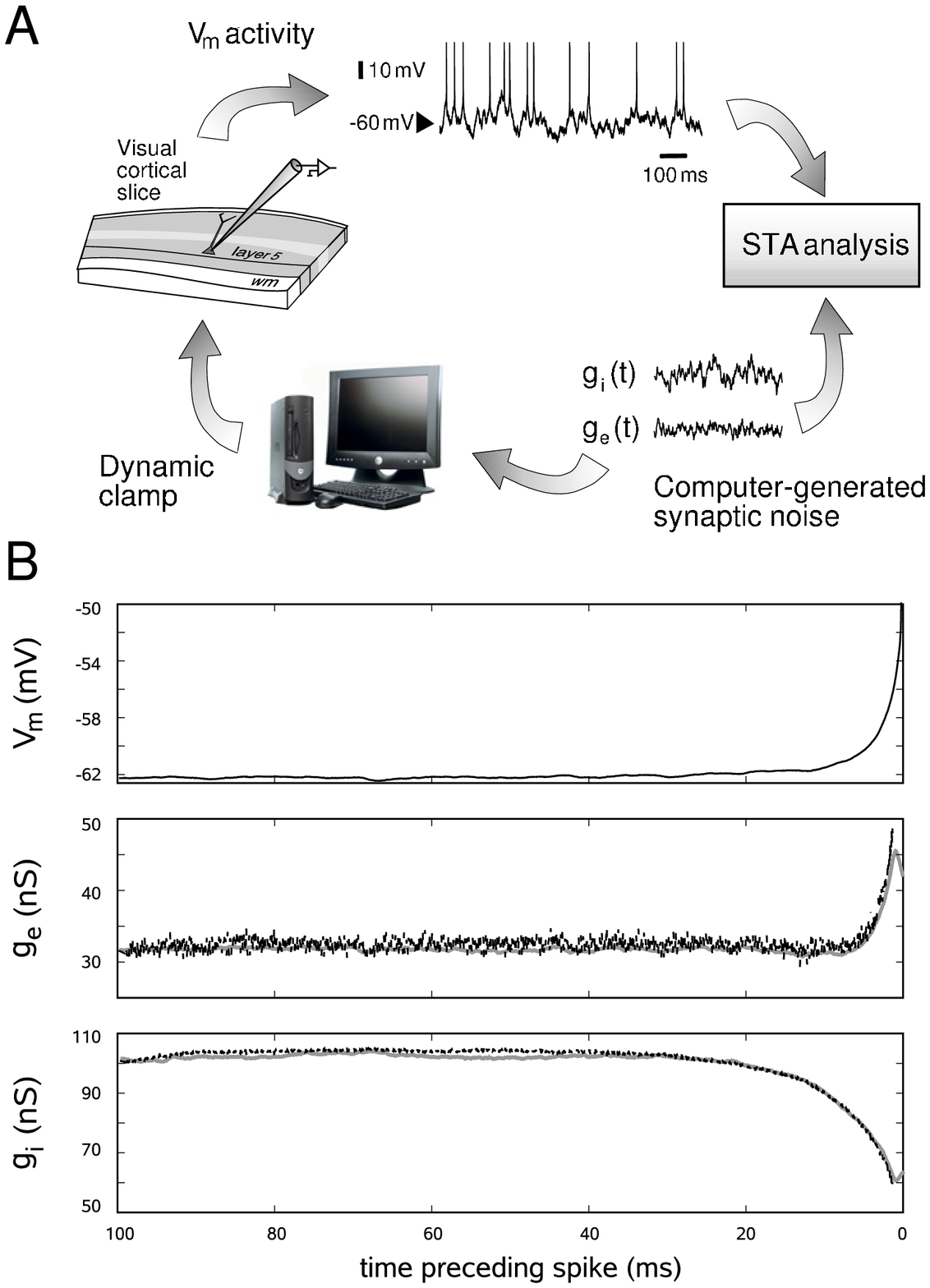}

    \caption{\small \figE}

    \label{fig:dc}
\end{figure}

\preprint{\noindent Figure~\ref{fig:dc}: \figE\\ }

\begin{figure}
    \includegraphics[width=14cm]{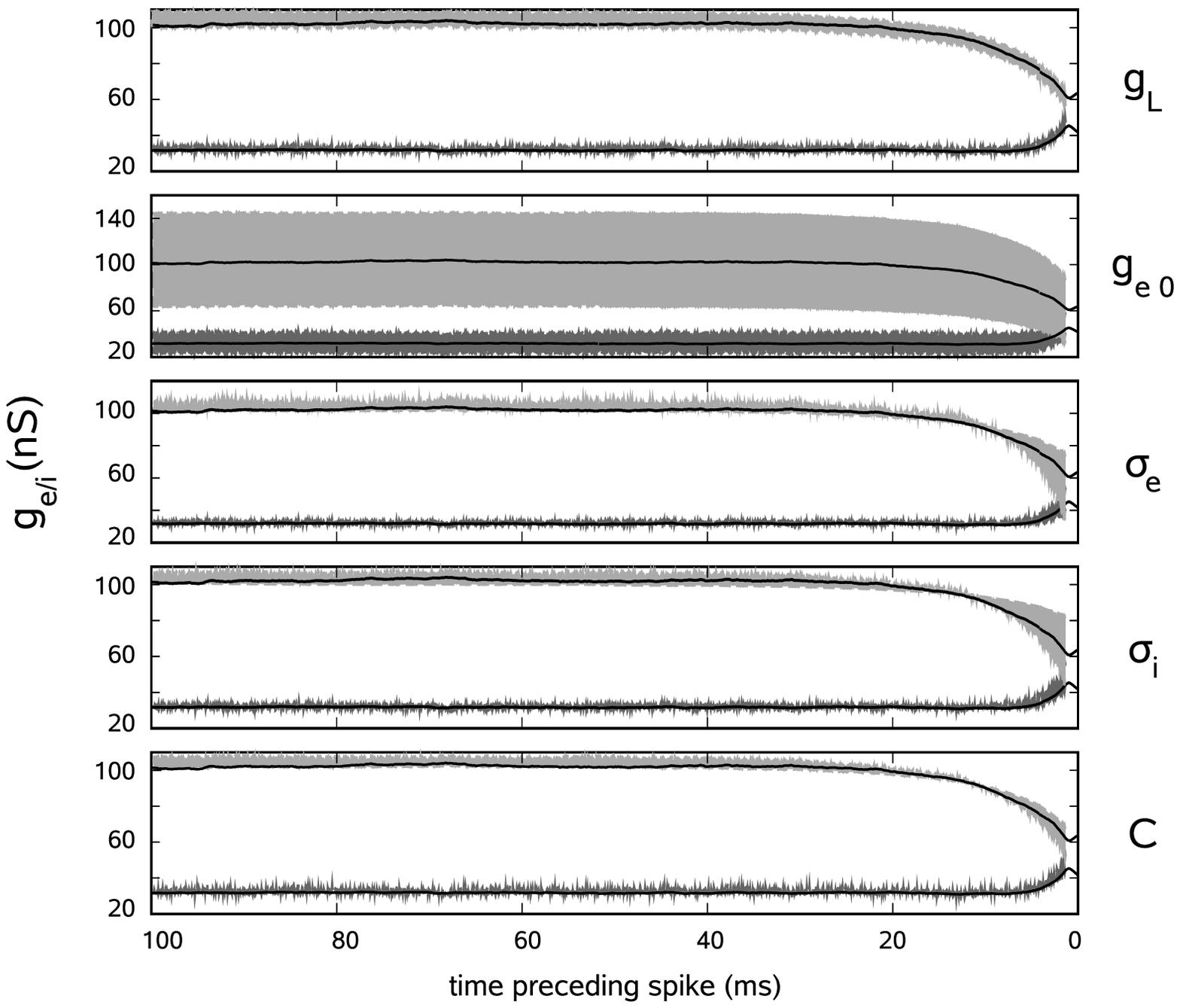}

    \caption{\small \figF}

    \label{fig:deviations}
\end{figure}

\preprint{\noindent Figure~\ref{fig:deviations}: \figF\\ }

\begin{figure}
    \includegraphics[width=14cm]{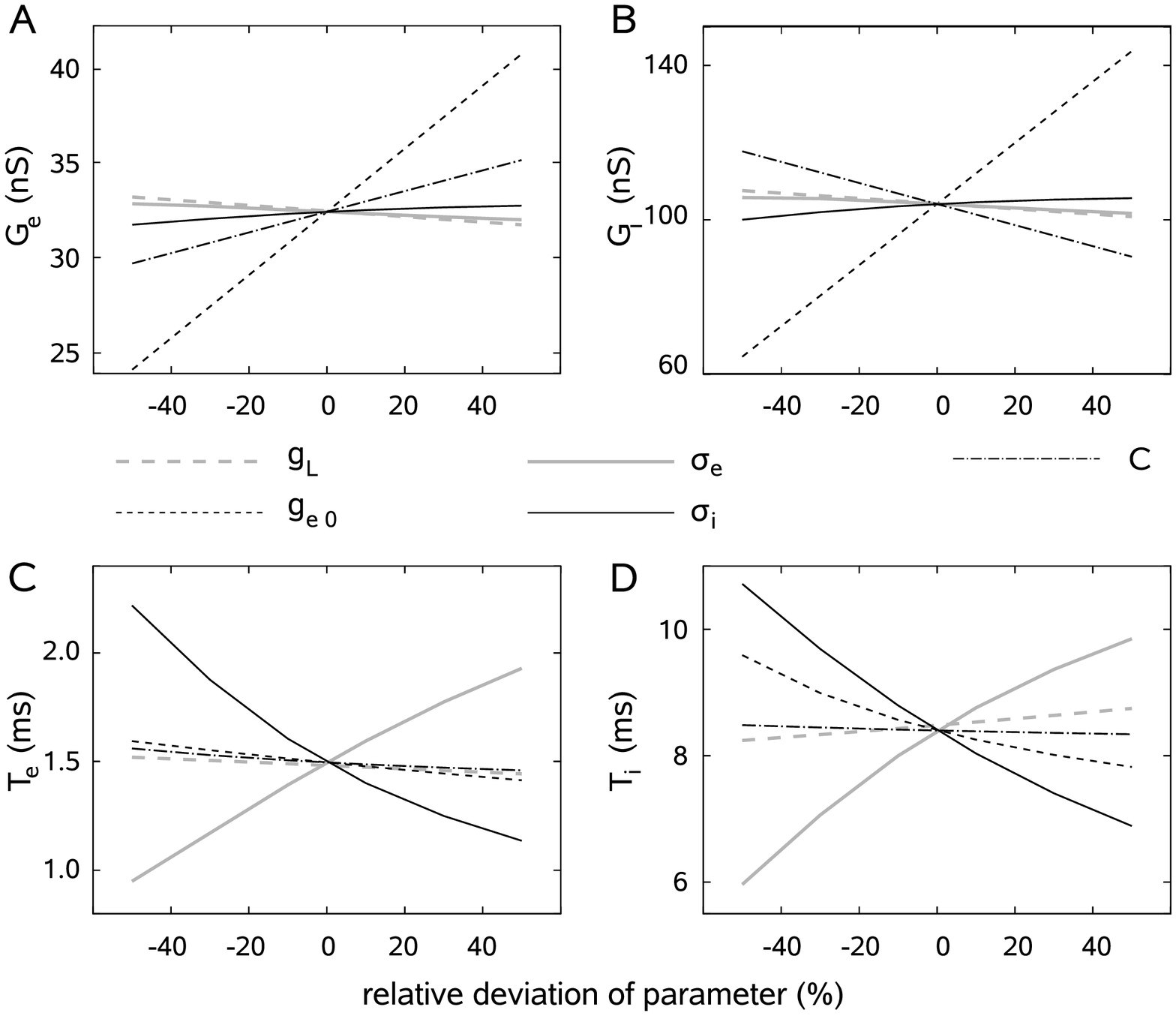}

    \caption{\small \figG}

    \label{fig:deviations_details}
\end{figure}

\preprint{\noindent Figure~\ref{fig:deviations_details}: \figG\\ }

\begin{figure}
    \includegraphics[width=10cm]{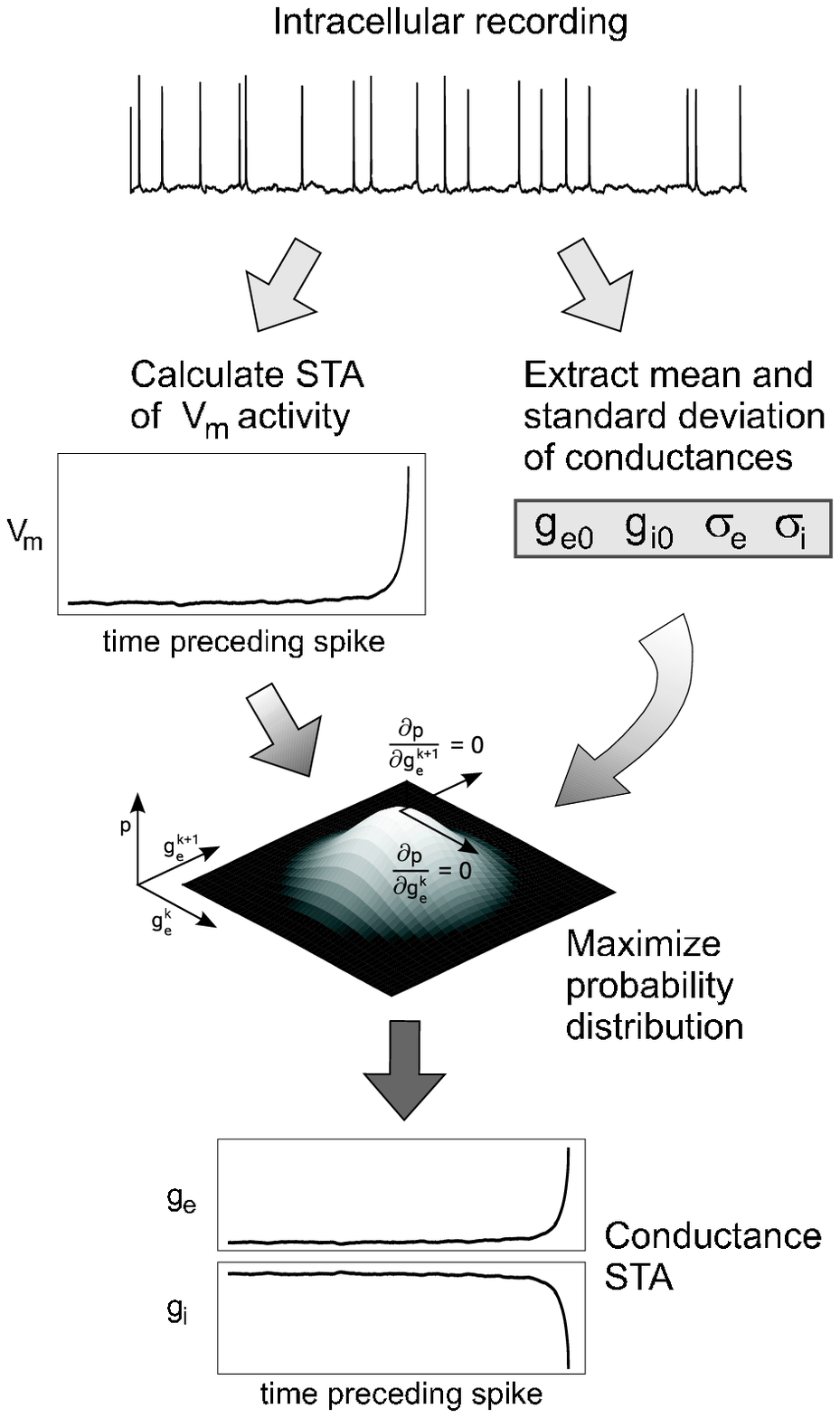}

    \caption{ \small \figH}

    \label{fig:sc}
\end{figure}

\preprint{\noindent Figure~\ref{fig:sc}: \figH}

%---------------------- FIGURES --------------------------------------

% insert EPS copies of the figures together with the legend in small

%----------------------------------------------------------------------------
%----------------------------------------------------------------------------

%----------------------------------------------------------------------------
%----------------------------------------------------------------------------

%----------------------------------------------------------------------------
%----------------------------------------------------------------------------

%----------------------------------------------------------------------------
%----------------------------------------------------------------------------

%----------------------------------------------------------------------------
%----------------------------------------------------------------------------

%----------------------------------------------------------------------------
%----------------------------------------------------------------------------

%----------------------------------------------------------------------------
%----------------------------------------------------------------------------

%----------------------------------------------------------------------------
%----------------------------------------------------------------------------

\end{document}